\documentclass[conference]{IEEEtran}
\IEEEoverridecommandlockouts

\usepackage[utf8]{inputenc}
\usepackage{ifthen}
\usepackage{hyperref}

\usepackage[dvipsnames]{xcolor}
\usepackage{soul}
\usepackage[normalem]{ulem}
\usepackage{graphicx}

\usepackage{epsf,picinpar}
\usepackage{varioref}
\usepackage{fdsymbol}

\usepackage{enumerate}
\usepackage{booktabs}

\usepackage{pifont}
\usepackage[pscoord]{eso-pic}

\usepackage{fancyvrb}

\usepackage{orcidlink}
\usepackage{float}
\usepackage{array}

\usepackage[color=red!50!white,textsize=footnotesize]{todonotes}

\usepackage{xurl}
\usepackage[framemethod=TikZ]{mdframed}

\usepackage[backend=biber,defernumbers=true,style=numeric-comp,maxcitenames=2,maxbibnames=3]{biblatex}

\usepackage[nomessages]{fp}%

\usepackage{subcaption}
\usepackage{wrapfig}
\usepackage{adjustbox}

\usepackage{lipsum}
\newcommand{\secref}[1]{Sec.~\ref{#1}}
\newcommand{\figref}[1]{Fig.~\ref{#1}}
\newcommand{\tabref}[1]{Tab.~\ref{#1}}

\newcommand{\xtodo}[1]{
    \ifthenelse{\boolean{showannotations}}%
    {\ifthenelse{\equal{#1}{}}{\textcolor{red}{TODO}}{\textcolor{red}{TODO:~{#1}}}}%
    {}%
}

\makeatletter
 \if@todonotes@disabled
 
 \else
 
 \fi
 \makeatother

\newcommand{\assignedto}[1]{%
    \ifthenelse{\boolean{showannotations}}%
    {\textbf{\noindent\ding{46}\textcolor{white}{~\colorbox{\assignementcolor}{Assigned to:}}~\textcolor{\assignementcolor}{#1}\\}%
    }
    {}
}

\makeatletter
\newcommand\notsotiny{\@setfontsize\notsotiny{7.5}{7.3}}
\makeatother

\newcommand{\rephrase}[1]{\textcolor{purple}{\uwave{#1}}} 

\renewcommand{\nb}[4]{
    \ifthenelse{\boolean{showannotations}}%
    {\fcolorbox{gray}{#2}{\bfseries\sffamily\scriptsize{#1}}
	{\sf\small$\blacktriangleright$\textcolor{#4}{\textit{#3}}$\blacktriangleleft$}}%
    {}%
}

\newcommand\personmarker[2]{\noindent\nb{#1}{yellow}{#2}{VioletRed}}
\newcommand\id[1]{\noindent\personmarker{ID}{#1}}

\newcommand{\rem}[1]{%
    \ifthenelse{\boolean{showannotations}}%
    {\textcolor{\oldtextcolor}{\st{#1}}}%
    {}%
}

\newcommand\add[1]{%
    \ifthenelse{\boolean{showannotations}}%
    {\textcolor{\newtextcolor}{{#1}}}%
    {#1}%
}

\newcommand\addblockbegin{%
    \ifthenelse{\boolean{showannotations}}%
    {\color{\newtextcolor}}%
    {}%
}

\newcommand\addblockend{%
    \ifthenelse{\boolean{showannotations}}%
    {\color{black}}%
    {}%
}

\newcommand\rep[2]{%
    \ifthenelse{\boolean{showannotations}}%
    {\rem{#1}~\add{#2}}%
    {#2}%
}

\newcommand{\placetextbox}[3]{
  \setbox0=\hbox{#3}
  \AddToShipoutPictureFG*{
    \put(\LenToUnit{#1\paperwidth},\LenToUnit{#2\paperheight}){\vtop{{\null}\makebox[0pt][c]{#3}}}%
  }%
}%

\newcommand{\corner}{$\rotatebox[origin=c]{180}{$\Lsh$}$}

\newcommand{\included}[1]{\textbf{$\rightarrow${#1}}}

\newenvironment{conclusionframe}[1]
  {\mdfsetup{
    frametitle={\colorbox{white}{\space#1\space}},
    innertopmargin=-1pt,
    frametitleaboveskip=-\ht\strutbox,
    frametitlealignment=\center
    }
  \begin{mdframed}[nobreak=true]%
  }
  {\end{mdframed}}

\newenvironment{recoframe}[1]
  {\mdfsetup{
    frametitle={\colorbox{white}{\space#1\space}},
    innertopmargin=-1pt,
    frametitleaboveskip=-\ht\strutbox,
    frametitlealignment=\center
    }
  \begin{mdframed}[nobreak=true]%
  }
  {\end{mdframed}}

\newcommand\xofy[3]{%
    \FPeval{perc}{round(100.0 / #2 * #1, 1)}%
    \ifthenelse{\equal{#3}{}}{%
        {#1} of {#2} (\perc\%)%
    }%
    {%
        {#1} of {#2} {#3} (\perc\%)%
    }%
}

\newcommand\xofyp[2]{%
    \FPeval{perc}{round(100.0 / #2 * #1, 1)}%
    {#1} of {#2} -- \perc\%%
}

\newcommand\perc[2]{%
    \FPeval{perc}{round(100.0 / #2 * #1, 1)}%
    \perc%
}

\newcommand\percp[3]{%
    \FPeval{perc}{round(100.0 / #2 * #1, 1)}%
    \ifthenelse{\equal{#3}{}}{%
        \perc\% ({#1} of {#2})%
    }%
    {%
        \perc\% ({#1} of {#2} {#3})%
    }%
}

\newcommand{\colheight}{-.25}
\newlength{\maxlen}

\newcommand{\numberofstudies}{24}

\newcommand{\maindatabar}[2][\numberofstudies]{%
    \databar[#1]{#2}{32}{barcolor}
}

\newcommand{\subdatabar}[2][\numberofstudies]{%
    \databar[#1]{#2}{36}{subbarcolor}
}

\newcommand{\subsubdatabar}[2][\numberofstudies]{%
    \databar[#1]{#2}{40}{subsubbarcolor}
}

\newcommand{\databar}[4][\numberofstudies]{%
    \settowidth{\maxlen}{#3}%
    \addtolength{\maxlen}{\tabcolsep}%
    \ifthenelse{\equal{#1}{}}{%
        \FPeval{perc}{round(100.0 / \studynumberdefault * #2, 1)}%
    }%
    {%
        \FPeval{perc}{round(100.0 / #1 * #2, 1)}%
    }%
    \FPeval\result{round(\perc / #3, 1)}%
    \rlap{\color{#4}\hspace*{-.5\tabcolsep}\rule[\colheight\ht\strutbox]{\result\maxlen}{1.2\ht\strutbox}}%
    \makebox[\dimexpr\maxlen-\tabcolsep][l]{#2 (\perc\%)}%
}

\newcommand{\subcategory}{\;\;\corner{}}
\newcommand{\subsubcategory}{\;\;\;\;\corner{}}

\newcommand{\quotetext}[1]{``\textit{#1}''}

\newcommand{\observation}[1]{%
    \noindent\textbf{Key observation: #1.}
}

\newcommand{\hreffinternal}[3]{\href{#1}{\textcolor{#3}{#2}}}

\newcommand{\hreff}[2]{\hreffinternal{#1}{#2}{linkblue}}
\newboolean{showannotations}
\setboolean{showannotations}{true} 

\newboolean{shownames}
\setboolean{shownames}{true}

\newcommand{\newtextcolor}{blue}
\newcommand{\oldtextcolor}{red}
\newcommand{\assignementcolor}{orange}
\definecolor{linkblue}{RGB}{120,106,237}
\definecolor{highlightcolor}{rgb}{.99, 1, .0}
\sethlcolor{highlightcolor}
\definecolor{barcolor}{HTML}{85d4ff}
\definecolor{subbarcolor}{HTML}{8fff85}
\definecolor{subsubbarcolor}{HTML}{a5bfcf}

\DeclareFieldFormat{labelnumberwidth}{#1\adddot}
\DeclareSourcemap{
  \maps[datatype=bibtex, overwrite]{
    \map{
      \perdatasource{bib/primary-studies.bib}
      \step[fieldset=KEYWORDS, fieldvalue=primary, append]
    }
  }
}

\begin{document}

\placetextbox{0.5}{0.99}{\colorbox{gray!3}{\textcolor{WildStrawberry}{Author pre-print. Publication accepted for \hreff{https://conf.researchr.org/home/ict4s-2025}{ICT4S 2025}. The final published version may differ.}}}%


\placetextbox{0.5}{0.05}{\small\colorbox{gray!3}{\textcolor{WildStrawberry}{Author pre-print. Publication accepted for \hreff{https://conf.researchr.org/home/ict4s-2025}{ICT4S 2025}. The final published version may differ.}}}%

\title{Bridging the Silos of Digitalization and Sustainability by Twin Transition:\\ A Multivocal Literature Review
\thanks{We acknowledge the support of the Natural Sciences and Engineering Research Council of Canada (NSERC), DGECR-2024-00293 (End-to-end Sustainable Systems Engineering).}}

\author{
     \IEEEauthorblockN{Baran Shajari\IEEEauthorrefmark{1}\orcidlink{0009-0005-3133-6441}, Istvan David\IEEEauthorrefmark{1},\IEEEauthorrefmark{2}\orcidlink{0000-0002-4870-8433}}
     \IEEEauthorblockA{\IEEEauthorrefmark{1}McMaster University, Canada -- \{shajarib, istvan.david\}@mcmaster.ca}
     \IEEEauthorblockA{\IEEEauthorrefmark{2}McMaster Centre for Software Certification (McSCert), Canada}
}

\maketitle

\begin{abstract}
Twin transition is the method of parallel digital and sustainability transitions in a mutually supporting way or, in common terms, ``greening of and by IT and data.'' Twin transition reacts to the growing problem of unsustainable digitalization, particularly in the ecological sense. Ignoring this problem will eventually limit the digital adeptness of society and the problem-solving capacity of humankind. Information systems engineering must find ways to support twin transition journeys through its substantial body of knowledge, methods, and techniques. To this end, we systematically survey the academic and gray literature on twin transition, clarify key concepts, and derive leads for researchers and practitioners to steer their innovation efforts.
\end{abstract}

\section{Introduction} \label{sec:introduction}

Digital transformation has become an essential tool for companies to improve their operational excellence~\cite{kraus2021digital}. Improved digitalization enables an array of competitive advantages, including enhanced data collection and management, quality improvements to products and services, and cost reduction~\cite{vial2019understanding}.

Unfortunately, the benefits of improved digitalization come at the price of increased environmental footprint~\cite{bianchini2023environmental}.
Information and Communications Technology (ICT) currently contributes to about 2-4\% of global CO$_2$ emissions---comparable to the carbon emissions of the avionics sector---and this number is projected to increase to about 14\% by 2040~\cite{belkhir2018assessing} due to computation-heavy digital enablers, such as AI and big data. This growth is unsustainable. To follow suit with the rest of the economy, the ICT sector should---directly or indirectly---decrease its CO$_2$ emissions by 42\% by 2030, 72\% by 2040, and 91\% by 2050~\cite{ituict}. 
Recently, companies have become more cognizant of the value of becoming environmentally sustainable, and the ways digitalization can aid such ambitions~\cite{aksin_sivrikaya2017where}. While digitalization exerts increasingly higher environmental pressure, it also opens opportunities in understanding, assessing, and enforcing sustainability imperatives, e.g., through targeted data collection and process optimization~\cite{bork2024role}. Sustainability and digitalization seem to be mutually dependent and inextricably linked~\cite{david2024circular}. This poses novel challenges for companies that strive to be competitive, but (environmentally and socially) responsible at the same time. Pursuing such joint innovation agendas requires novel methods to strike a balance between green and digital transitions.

Originally suggested in the European Green Deal~\cite{eucommission2022twin}, twin transition is the paradigm of ``\textit{greening of and by IT\& data}''~\cite{blum2022what}, i.e., fostering reinforcing relationships between digital and sustainability transitions.
As such, twin transition helps \textbf{bridge the silos of digitalization and sustainability}.
While its benefits are clear, twin transition is not well-understood, and supporting methods are in their infancy.

\subsection*{Contributions}

This work is the first to systematically study the topic of twin transition in scientific and gray literature.

Motivated by the early stage of research and limited academic literature, we opt for a multivocal study~\cite{garousi2019guidelines}, i.e., we include non-academic (``gray'') literature in our study, e.g., pre-prints and news articles.\footnote{Replication package: \url{https://zenodo.org/records/15258629}.}
Our findings highlight the misalignment of the necessary technological advancement with the mostly non-technical stakeholders. We believe that the information systems community has a lot to offer in the development of twin transition methods and tools.

\section{Background}\label{sec:background}


\subsection{Digital transformation}

Digital transformation is the process of organizations adapting to changes in their business through utilizing digital technology~\cite{vial2019understanding}.
As companies progress through their digital transformation journeys, they adjust their value-creation processes and strategies, largely driven by the opportunities in digitalization~\cite{kraus2021digital}.
Some of the key benefits of digital transformation include improved operational performance through higher degrees of automation, more effective information flow across corporate hierarchies, and improved decision-making~\cite{vial2019understanding}.

Digital transformation has been a topic of particular interest in the past decades~\cite{verina2019digital}.
First explored in the late 80s, the concept of digital transformation came to be when researchers studied how information technology (IT) impacts organizational structures, performance, and innovation~\cite{plekhanov2023digital}. With the expansion of the scope of IT systems, digital transformation research has increasingly incorporated a wider range of disciplines within management, business, and economics.
Today, we see an increasing interest in aligning digital transformation principles with corporate sustainability ambitions~\cite{lazazzara2024digital}.



\subsection{Sustainability}

A commonly used view of sustainability originates from \textcite{brundtland1987our} who defines sustainability as
the capacity to ``\textit{meet the needs of the present without compromising the ability of future generations to meet their own needs}.'' This ambition is refined into three dimensions: economic (financial viability), environmental (reduction of ecological impact), and societal sustainability (increased social utility).
\textcite{hilty2006relevance} provides a more technologically attuned definition as the ability to ``\textit{preserve the function of a system over an extended period of time}.'' To harmonize technology and sustainability, \textcite{penzenstadler2013generic} recommend technical sustainability as the fourth dimension, commonly associated with the longevity of systems and their ability to adapt.

Sustainability has become a key value driver~\cite{chiu2020can}, motivating companies to incorporate its various forms in their value propositions and operations~\cite{david2025susdevops}. These efforts lead to ecologically and economically friendly practices, fostering circularity~\cite{david2024circular} and value retention across value chains~\cite{stucki2024data}.

Digitalization is a key enabler of sustainability~\cite{bork2024role}. However, the sustainability of digitalization must be considered, too~\cite{david2024on}. We focus on the mutually reinforcing effects of digital and sustainability transitions---i.e., twin transition~\cite{blum2022what}.


\subsection{Related work}

%
Through a scoping review of 112 contributions, \textcite{mouthaan2023systemic} find that research on digitalization and sustainability transition tends to focus on digital technologies, overlooking the complex and uncertain effects of digital technologies on sustainability.
A scientometric study by \textcite{wang2020making} corroborates these findings and shows that concepts of sustainability, sustainable development, energy, and governance play a connecting role between the two fields. This suggests that more research from the green transformation angle might have the most impact on how industries advance.
Alas, bibliometric studies suggest that the number of academic works on digital transformation is five times higher than sustainability transition~\cite{burinskiene2024digital}.
\textcite{ortega-gras2021twin} provide recommendations for successful twin transition in an Industry 4.0 context---albeit at a high strategic level---including clear digitalization and circularity objectives, and employee upskilling.
To support firms in planning for simultaneous digital and green transitions, \textcite{pan2023knowledge} define a taxonomy for transition design.

\section{Study design}\label{sec:study-design}



The goal of this study is to identify and analyze the characteristics, components, concerns, and stakeholders of twin transition (TT) from a researcher’s viewpoint. To meet our goal, we formulate the following research questions.

\begin{enumerate}[\bfseries{RQ}1.]
    \item \textbf{What are the various \ul{definitions} of TT?}

    We aim to \textbf{consolidate} the various notions of twin transition, understand its main concerns and components, and their relationship.

    \item \textbf{What are the \ul{sustainability ambitions} of TT?}
    
    We \textbf{identify} the sustainability dimensions and their combinations in TT.

    \item \textbf{Who are the \ul{stakeholders} and \ul{users} involved in TT?}

    We \textbf{identify} roles, units, organizations who articulate ambitions of twin transition,
    as well as roles involved in implementing twin transition.



    \item \textbf{What are the \ul{requirements} and \ul{enablers} of TT?}

    We identify technical and business capabilities, resources, methods, and tools that \textbf{positively impact} the success of twin transition.
    
    \item \textbf{What are the \ul{challenges} that hinder TT?}

    We identify phenomena that \textbf{adversely impact} the success of TT.
    

\end{enumerate}

\subsection{Methodology: Multivocal study}

To answer the research questions, we
conduct a \textit{multivocal literature review} (MLR)~\cite{garousi2019guidelines}. An MLR includes a systematic review of both academic and so called ``gray'' literature (GL). Examples of gray studies include, pre-prints, reports, white papers, news articles, etc.
MLRs are gaining popularity in computer science
and their added value has been recognized in management and organizational studies as well~\cite{adams2017shades}.
We follow \textcite{garousi2019guidelines} in the design of our study.

\subsubsection{Assessing the need for an MLR}

To assess the need for an MLR, we first informally survey the topic of twin transition to understand its publication dynamics, and subsequently, we apply the checklist of \textcite[Tab 4]{garousi2019guidelines}.
%
The checklist
stipulates answering at least one of its seven questions positively to suggest the inclusion of GL. We answer (at least) two questions positively: there is a large volume of practitioner sources (Q7 in \cite{garousi2019guidelines}) and the subject can be investigated only partially through formal literature due to the early stage of research (Q1 in \cite{garousi2019guidelines}). We conclude that the inclusion of GL is justified.
%
We target Tier-1 GL (i.e., high outlet control and credibility: theses, government reports, white papers, etc.) with some instances of Tier-2 GL allowed (moderate outlet control and credibility: annual reports, news articles, etc.)~\cite{adams2017shades}.

\subsection{Search and selection}

\subsubsection{Databases}
To search for potentially relevant studies, we use the following sources. Indexing sites: Scopus, Web of Science; Computer Science and Software: ACM Digital Library, IEEE Xplore; Business and management: EBSCO, Business Source Complete, ProQuest (including ABI/Inform); preprint sites: arXiv.org, Scopus preprints; mixed: Google Scholar. ProQuest, and Business Source Complete index both academic and gray literature. The content on arXiv.org is gray literature in general, and often, preprints of eventual academic publications. Scopus indexes preprints separately, back until 2017, which is sufficient for our purposes. We round out the search with a query on Google Scholar. Due to the limitations of Google Scholar~\cite{bonato2016google}, we search in abstracts only for ``twin transition'' and by ordering articles by date.


\subsubsection{Search string}
We construct the search string from the \textcolor{VioletRed}{exact search term} and its refined form consisting of the \textcolor{ForestGreen}{sustainability} and \textcolor{blue}{digitalization} components. (Database-specific search strings are available in the replication package.)

\phantom{}

\begin{footnotesize}
\begin{Verbatim}[commandchars=\\\{\}]
(\textcolor{VioletRed}{"twin transition"}) OR
(
 (\textcolor{ForestGreen}{"sustainability transition" OR "green transition"})
  AND
 (\textcolor{blue}{"digital transition" OR "digital transformation"})
)
\end{Verbatim}
\end{footnotesize}

\phantom{}

\subsubsection{Automated search}

We query the databases on September 3, 2024. We search in the title, abstract, and keywords of papers in most databases. On arXiv, we search in the title, abstract, and comments.
We retrieve a total of \textbf{781 references}. Details are reported in \tabref{tab:numbers}.
We remove 309 duplicates and arrive at 472 unique references that undergo selection.


\begin{table}[t]
\renewcommand{\arraystretch}{.75}
\centering
\caption{Statistics: search and selection}
\label{tab:numbers}
\begin{tabular}{@{}l@{\hspace{1em}}rrrr@{}}
\toprule
\multicolumn{1}{c}{\textbf{}} & \multicolumn{1}{c}{\textbf{All}} & \multicolumn{1}{c}{~~\textbf{Excl}} & \multicolumn{1}{c}{\textbf{Incl}} & \multicolumn{1}{c}{\textbf{$\kappa$}} \\ \midrule
Duplicate/non-English & 781 & 309 & 472 & \\
E0 &  & 250 & 222 & \\
\midrule
E1 & & 11 & & \\
E2 & & 2 && \\
E3 (Parallel transition) & & 66 & \\
E4 (Informed transition) & & 71 & \\
E5 & & 48 & \\
 &  & 198 & ~~\textbf{24} (\textbf{3.07\%}) & ~~0.854 \\
\bottomrule
\end{tabular}
\end{table}

\subsubsection{Selection}

We exclude unrelated references by the following exclusion criteria.
%
\begin{enumerate}[\bfseries{E}1.]
\setcounter{enumi}{-1}
    \item Not accessible (not English, cannot download); different notion of twin transition (e.g., chemistry); not primary research (e.g., reviews); other irrelevant artifacts (datasets, forewords and editorials, full proceedings, grants).
    
    \item Digitalization aspect is missing or unclear. 
    
    \item Sustainability aspect is missing or unclear. 
    
    \item Digitalization is clear but not related to sustainability. 
    
    \item Sustainability is clear but not related to digitalization. 
    
    \item Unrelated for other reasons.
\end{enumerate}
A primary study is excluded if it meets at least one exclusion criterion.
E0 is straightforward to evaluate, thus, one author evaluates it and the other author validates the decisions.
For exclusion criteria E1--E5, each primary study is evaluated by both authors independently.
In case of a tie, discussion is facilitated. We observe a particularly \textbf{high Cohen-$\kappa$} of 0.854, i.e., almost perfect agreement.
Eventually, we include \textbf{24 primary studies} (\tabref{tab:numbers}) as \textit{twin transition} studies. We answer the RQs using these 24 studies. In RQ1, we will also consider excluded papers that label their approach as twin transition, but fail to meet its criteria. These are \textbf{E3} and \textbf{E4}, and as explained in \secref{sec:results-rq1}, we label them as \textit{parallel transition} and \textit{informed transition}, respectively.

\subsection{Threats to validity and study quality}


\paragraph{Construct validity}
Our observations are artifacts of the sampled studies and selection bias may threaten the construct validity of this study. To mitigate this threat, both authors inspect the studies per the recommendations of \textcite{kitchenham2004procedures}. Another threat arises from the early stage of research and limited body of knowledge on twin transition. To mitigate this threat, we opt for an MLR.
\paragraph{Internal validity}
The term ``twin transition'' is used with different meanings in some natural sciences. This leads to a number of false positives in our search.
To mitigate this threat, we inspect every abstract and publication venues' disciplines thoroughly.
%
%
Another threat to internal validity stems from the manual classification of topics in RQs 2--5.
To mitigate this threat, we facilitate in-depth discussions to arrive at the eventual classifications after a consensus.
%
%
\paragraph{External validity}
Our takeaways are valid for TT and any generalization must be approached cautiously.
Specifically, our takeaways may not be valid within the confines of digital transformation without sustainability considerations, and sustainability transition without digitalization considerations.
%
%
\paragraph{Study quality.} Our work scores \percp{7}{11}{points} in the quality checklist of \textcite{petersen2015guidelines}. (Need: 1 point; search: 1; evaluation of the search: 2; extraction and classification: 2; validity: 1.) This quality score \textit{significantly} exceeds the scores reported from software engineering---33\% median, with only 25\% of studies above 40\%~\cite{petersen2015guidelines}.
Thus, we consider our study of \textbf{high quality}.


\subsection{Publication trends}

\begin{figure}[t]
    \includegraphics[width=\linewidth]{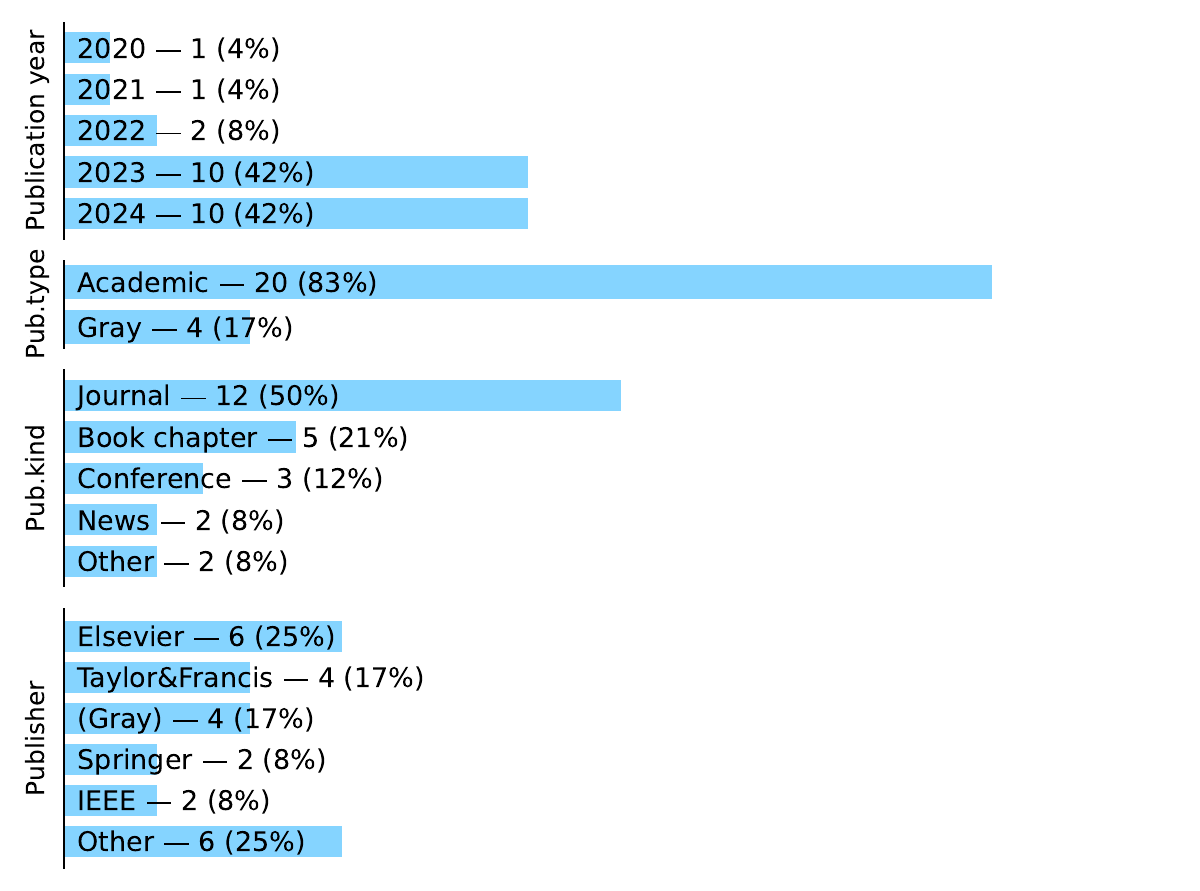}
    \caption{Studies on twin transition (as of September 2024)}
    \label{fig:pubstats}
\end{figure}


The number of publications (\figref{fig:pubstats}) shows an increasing trend, with the past two years constituting 84\% of the corpus. 2024 is a partial year in our work (studies published until September), i.e., the increasing trend in publications is expected to continue.
83\% of the corpus consists of academic studies, with 17\% grey studies rounding out our sample. 71\% of the studies are journal articles and book chapters, suggesting mature research to draw on.
%
%
We judge the corpus to be in a good shape to allow for sound conclusions.

\section{Results}\label{sec:results}

\begin{figure}[t]
    \centering
    \includegraphics[width=0.75\linewidth]{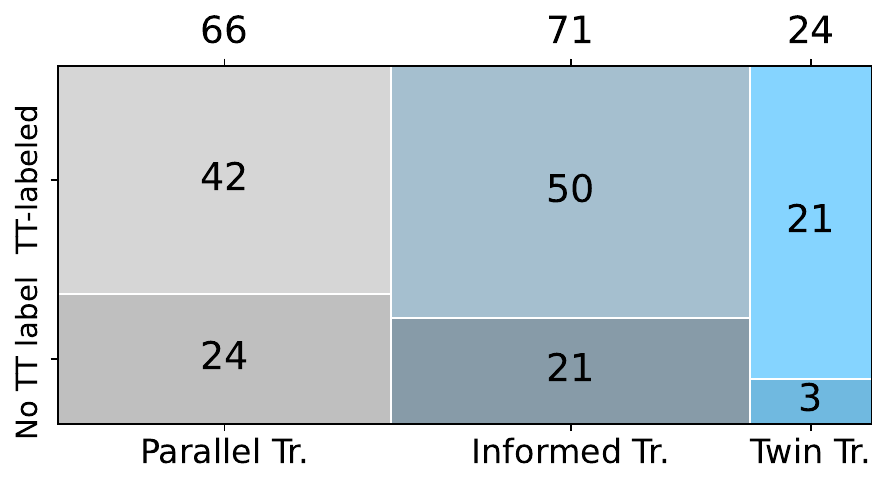}
    \caption{Proportions of various digital/green transition concepts}
    \label{fig:tt-notions}
\end{figure}

\begin{table}[h]
\centering
\caption{Twin transition works (not) labeled labeled as such}
\label{tab:results-rq1}
{\footnotesize
\begin{tabular}{@{}llp{4cm}@{}}
\toprule
\multicolumn{1}{c}{\textbf{Domain}} & \multicolumn{1}{c}{\textbf{\#Studies}} & \multicolumn{1}{c}{\textbf{Studies}} \\ \midrule

\textbf{Labeled as TT} & \maindatabar{21} & \\

\subcategory{} Academic & \subdatabar{17} & \cite{balanici2023f5g, bianchini2023environmental, brueck2024chinas, chen2023enabling, gao2024eus, gigauri2023digital, korucuk2022assessing, kovacic2024twin, makitie2023digital, matinmikko-blue2021sustainability, meijer2024perspectives, muller2023digitisation, niet2024framing, poscic2024role, rehman2023twin, sanchez2024european, stucki2024data} \\
\subcategory{} Grey & \subdatabar{4} & \cite{baum2024will, celeste2023digital, myyrylainen2023small, pan2024policy}\\

\textbf{Not labeled as TT} & \maindatabar{3} & \\ \subcategory{} Academic & \subdatabar{3} & \cite{chiu2020can, reich2024towards, tekavc2023pilot} \\

\bottomrule
\end{tabular}}
\end{table}

\subsection{Definitions of twin transition (RQ1)}\label{sec:results-rq1}

\newcommand{\allTransitions}{161}
\newcommand{\ttLabel}{113}
\newcommand{\pTransition}{66}
\newcommand{\iTransition}{71}
\newcommand{\tTransition}{24}

To consolidate the various notions of twin transition, we analyze the \textit{full corpus} of 161 studies and identify studies that focus on both digital and sustainability transitions. These are the studies that we identified as the final set of primary studies (\tTransition{} studies) and, in addition, those that we excluded from the final set on account of missing or one-directional link between digital and sustainability transitions (E3 and E4 in \tabref{tab:numbers}; \pTransition{} and \iTransition{} studies, respectively). Thus, in total, there are \allTransitions{} studies that focus on digital and sustainability transition, but the link between the two transitions varies.
%
%
We distinguish between three transition models by the degree of coordination between the two transitions, as reported in below and in \figref{fig:tt-notions}.

\paragraph{Parallel transition} \xofy{\pTransition}{\allTransitions}{} studies do not recognize the link between the two transitions. In essence, digital and sustainability transitions happen without much synchronization or information exchange. We call this approach \textbf{parallel transition}. \xofy{42}{\pTransition}{} studies label themselves as twin transition; alas, they fall short of true twin transition due to the lack of coordination between the two transitions.

\paragraph{Informed transition} \xofy{\iTransition}{\allTransitions}{} studies recognize the link between the two transitions, but only in one direction. In all \iTransition{} cases, it is digitalization that serves sustainability ambitions; while sustainability of digitalization is overlooked. We call this approach \textbf{informed transition} as some information is exchanged between the two transitions, although not enough to render the approach twin transition. \xofy{50}{\iTransition}{} informed transition works label themselves as twin transition, despite not meeting the necessary criterion of bi-directional informedness between the two transition threads.

\paragraph{Twin transition} \xofy{\tTransition}{\allTransitions}{} studies recognize the link between transitions in both directions, constituting the class of true \textbf{twin transition}. \xofy{21}{\tTransition}{} studies label themselves correctly; and additional \xofy{3}{\tTransition}{} achieve twin transition without labeling themselves as such.

\begin{conclusionframe}{RQ1: Definition of twin transition}
The vast majority, \percp{92}{\ttLabel}{} of the state of the art uses an imprecise notion of twin transition, in which digital and sustainability transitions are not, or only partially coordinated.
\end{conclusionframe}

\begin{figure}[h]
    \centering
    \includegraphics[width=\linewidth]{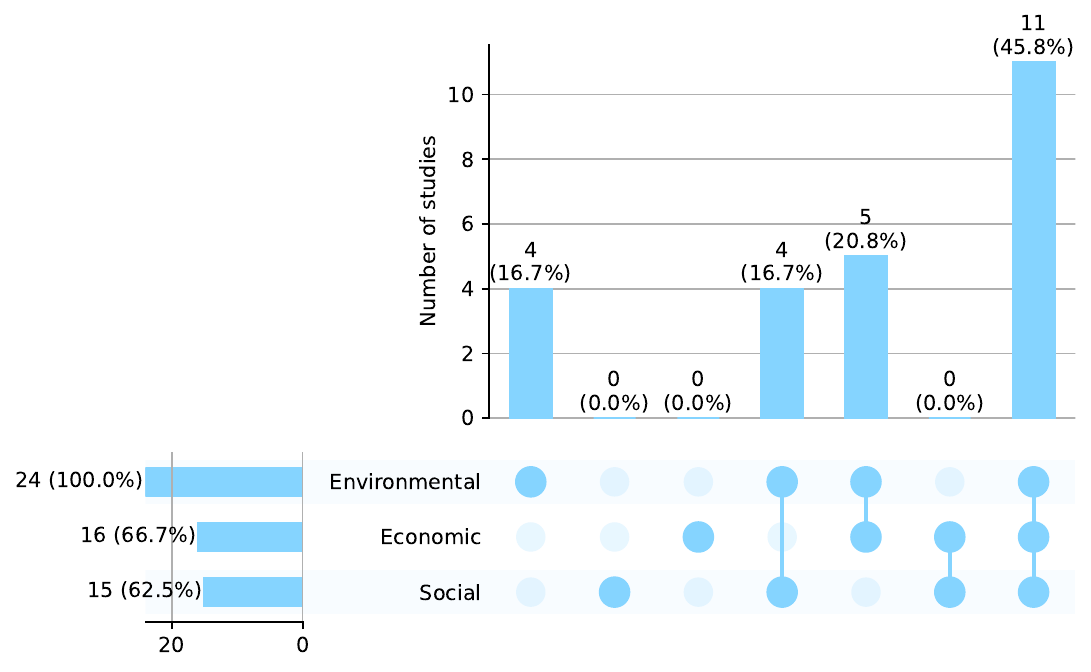}
    \caption{Breakdown of joint sustainability dimensions}
    \label{fig:rq2-sustainability-dims-upset}
\end{figure}

\subsection{Sustainability ambitions of twin transition (RQ2)}\label{sec:results-rq2}

\paragraph{Sustainability dimensions.} As shown in \figref{fig:rq2-sustainability-dims-upset}, we frequently encounter the three \textcite{brundtland1987our} sustainability dimensions in our sample. As expected, \textit{environmental} sustainability is a well-articulated ambition in \xofy{24}{24}{} studies, in line with the key points of the European Green Deal where twin transition originates from. \textit{Economic} and \textit{social} sustainability are well-represented in our sample as well, present in \xofy{16}{24}{} and \xofy{15}{24}{} studies, respectively. Interestingly, we do not find any of the two more recently acknowledged sustainability dimensions, technical~\cite{penzenstadler2013generic} and individual~\cite{duboc2019do} sustainability.

\paragraph{Combinations of sustainability dimensions.}

We observe a high number of studies that combine two or more sustainability dimensions. The mean number of sustainability dimensions is 2.29, and the mode is 3. It is only environmental sustainability that appears in isolation (\xofyp{4}{24}{}). In the vast majority of the sampled studies, in \xofy{20}{24}{}, environmental sustainability is coupled with at least one other sustainability dimension: in \xofy{4}{24}{studies} with social, in \xofy{5}{24}{studies} with economic, and in \xofy{11}{24}{studies} with both social and environmental sustainability.
The latter category of papers highlights the intricate links of sustainability dimensions, including social even in technical problems, such as low-carbon energy transitions~\cite{muller2023digitisation} and 6G spectrum management~\cite{matinmikko-blue2021sustainability}.
These observations are in line with, e.g., those of \textcite{celeste2023digital} who conclude that achieving a digital and just society \quotetext{cannot prescind from environmental considerations}.

\begin{conclusionframe}{RQ2: Sustainability ambitions}
{Twin transition inextricably links digitalization and environmental sustainability; at the same time, it also acknowledges (but does not emphasize) social and economic sustainability. Twin transition tends to focus on multiple sustainability dimensions at the same time, challenging their understanding and assessment.}
\end{conclusionframe}
\subsection{Stakeholders and user groups (RQ3)}

As shown in \tabref{tab:results-rq3-stakeholder-types}, non-technical stakeholders (e.g., regulators, policy-makers) are reported in \xofy{24}{24}{} studies, while technical stakeholders (e.g., IT leaders, technology vendors) are considered only in \xofy{4}{24}{}. We observe the unexpected lack of information systems experts, software engineers, and researchers of technical fields.

\begin{table}[h]
\footnotesize
\centering
\caption{Stakeholder types}
\label{tab:results-rq3-stakeholder-types}
\begin{tabular}{@{}lll@{}}
\toprule
\multicolumn{1}{c}{\textbf{Domain}} & \multicolumn{1}{c}{\textbf{\#Studies}} & \multicolumn{1}{c}{\textbf{Studies}} \\ \midrule

\textbf{Non-technical} & \maindatabar{24} & \cite{balanici2023f5g, baum2024will, bianchini2023environmental, brueck2024chinas, celeste2023digital, chen2023enabling, chiu2020can, gao2024eus, gigauri2023digital, korucuk2022assessing, kovacic2024twin, makitie2023digital, matinmikko-blue2021sustainability, meijer2024perspectives, muller2023digitisation, myyrylainen2023small, niet2024framing, pan2024policy, poscic2024role, rehman2023twin, reich2024towards, sanchez2024european, stucki2024data, tekavc2023pilot} \\

\textbf{Technical} & \maindatabar{4} & \cite{baum2024will,celeste2023digital,matinmikko-blue2021sustainability,muller2023digitisation} \\

\bottomrule


\end{tabular}
\end{table}


\tabref{tab:results-rq3-stakeholder-sectors} reports the primary sectors and affiliations of these stakeholder groups. We observe a particularly high frequency of \textit{government}-affiliated stakeholders (\xofyp{18}{24}), followed by a near-even distribution of stakeholders over \textit{businesses} (\xofyp{9}{24}), \textit{suppliers} (\xofyp{8}{24}), \textit{researchers and universities} (\xofyp{8}{24}), and \textit{corporate leadership} (\xofyp{6}{24}).

\begin{table}[t]
\footnotesize
\centering
\caption{Stakeholder sectors and affiliations}
\label{tab:results-rq3-stakeholder-sectors}
\begin{tabular}{@{}llp{4cm}@{}}
\toprule
\multicolumn{1}{c}{\textbf{Domain}} & \multicolumn{1}{c}{\textbf{\#Studies}} & \multicolumn{1}{c}{\textbf{Studies}} \\ \midrule

\textbf{Government} & \maindatabar{18} & \cite{baum2024will, bianchini2023environmental, brueck2024chinas, celeste2023digital, chiu2020can, gao2024eus, gigauri2023digital, korucuk2022assessing, kovacic2024twin, makitie2023digital, matinmikko-blue2021sustainability, meijer2024perspectives, myyrylainen2023small, niet2024framing, pan2024policy, poscic2024role, rehman2023twin, sanchez2024european} \\

\textbf{Businesses, companies} & \maindatabar{9} & \cite{celeste2023digital, chen2023enabling, chiu2020can, gao2024eus, kovacic2024twin, myyrylainen2023small, niet2024framing, poscic2024role, reich2024towards} \\

\textbf{Supply chain, logistics} & \maindatabar{8} & \cite{balanici2023f5g, baum2024will, celeste2023digital, chen2023enabling, korucuk2022assessing, makitie2023digital, muller2023digitisation, reich2024towards} \\

\textbf{Universities} & \maindatabar{8} & \cite{brueck2024chinas, celeste2023digital, makitie2023digital, matinmikko-blue2021sustainability, niet2024framing, pan2024policy, rehman2023twin, tekavc2023pilot}\\

\textbf{Corporations \& units} & \maindatabar{6} & \cite{baum2024will, korucuk2022assessing, kovacic2024twin, makitie2023digital, myyrylainen2023small, rehman2023twin} \\

\bottomrule


\end{tabular}
\end{table}

\begin{conclusionframe}{RQ3: Stakeholders and user groups}
The key stakeholders in twin transition lack a technical background and the vast majority are government-affiliated or business actors.
\end{conclusionframe}

\subsection{Enablers and requirements of twin transition (RQ4)}

\tabref{tab:results-rq4} reports the enablers and requirements of twin transition as reported in our sample. We observe that \textit{technical} enablers slightly outweigh \textit{non-technical} ones, encountered in \xofy{14}{24}{} and \xofy{12}{24}{} studies, respectively.

\paragraph{Technical enablers.}
Among the technical enablers, we observe the emergence of four major themes: specific \textit{technologies} (\xofyp{8}{24}{}), \textit{techniques and methods} (\xofyp{6}{24}{}), \textit{infrastructure} (\xofyp{5}{24}{}), and \textit{information systems} (\xofyp{3}{24}{}). AI is the most frequently encountered technological enabler. Authors often cite the role of AI in enabling smart cities~\cite{celeste2023digital} and the circular economy~\cite{gigauri2023digital}. At the same time, authors call for the alignment of AI with social values~\cite{pan2024policy} and for more transparent assessment of its ecological footprint~\cite{niet2024framing}.
\textit{Modeling and simulation} is identified as an enabling technique, e.g., for forecasting the carbon effects of digitalization~\cite{sanchez2024european}; while calling for actionable numeric methods over qualitative ones~\cite{stucki2024data}.

\paragraph{Non-technical enablers.}
Among the non-technical enablers, two topics are particularly often-discussed: \textit{policies and government}, found in \xofy{6}{24}{} studies; and \textit{innovation}-related concerns, found in \xofy{5}{24}{} studies. Some examples of enabling policies are strategic government actions to promote twin transition~\cite{korucuk2022assessing} and promoting voluntary codes of conduct, e.g., in data centers to reduce their energy consumption~\cite{celeste2023digital}.
Some studies call for stronger forms of innovation instead of incremental approaches, with explicit ambitions to \quotetext{reconfigure the entire structure of socio-technical systems to align with sustainability}~\cite{makitie2023digital} through novel design methods to embed sustainability into product and services development~\cite{chiu2020can}, and collaborative business models to involve political decision-makers~\cite{stucki2024data}.
%
%
Education of future workforce~\cite{tekavc2023pilot} and upskilling of the current~\cite{korucuk2022assessing} are identified as important enablers as well.

\begin{table}[t]
\footnotesize
\centering
\caption{Key enablers of twin transition}
\label{tab:results-rq4}
\begin{tabular}{@{}lll@{}}
\toprule
\multicolumn{1}{c}{\textbf{Domain}} & \multicolumn{1}{c}{\textbf{\#Studies}} & \multicolumn{1}{c}{\textbf{Studies}} \\ \midrule

\textbf{Technical} & \maindatabar{14} & \\

\subcategory{} Technology & \subdatabar{8} & \\
\subsubcategory{} \textit{AI} & \subsubdatabar{4} & \cite{celeste2023digital,gigauri2023digital,niet2024framing,pan2024policy} \\
\subsubcategory{} \textit{Big data} & \subsubdatabar{2} & \cite{chen2023enabling,gao2024eus} \\
\subsubcategory{} \textit{Other} & \subsubdatabar{3} & \cite{gigauri2023digital,makitie2023digital,reich2024towards} \\

\subcategory{} Techniques and methods & \subdatabar{6} & \\
\subsubcategory{} \textit{Real-time data and monitoring} & \subsubdatabar{5} & \cite{chen2023enabling,gigauri2023digital,makitie2023digital,muller2023digitisation,sanchez2024european} \\
\subsubcategory{} \textit{Modeling and simulation} & \subsubdatabar{2} & \cite{sanchez2024european,stucki2024data} \\
\subsubcategory{} \textit{Virtualization} & \subsubdatabar{1} & \cite{sanchez2024european} \\

\subcategory{} Infrastructure & \subdatabar{5} & \cite{balanici2023f5g,brueck2024chinas,celeste2023digital,matinmikko-blue2021sustainability,niet2024framing} \\

\subcategory{} Information systems & \subdatabar{3} & \cite{pan2024policy,reich2024towards,stucki2024data} \\

\textbf{Non-technical} & \maindatabar{12} & \\

\subcategory{} Policies and government & \subdatabar{6} & \cite{bianchini2023environmental,brueck2024chinas,celeste2023digital,korucuk2022assessing,meijer2024perspectives,pan2024policy} \\

\subcategory{} Innovation and env. mgmt & \subdatabar{5} & \cite{chen2023enabling,chiu2020can,makitie2023digital,rehman2023twin,stucki2024data} \\
\subcategory{} Education and upskilling & \subdatabar{2} & \cite{korucuk2022assessing,tekavc2023pilot}\\
\subcategory{} Customer behavior & \subdatabar{2} & \cite{chiu2020can,pan2024policy}\\

\bottomrule


\end{tabular}
\end{table}

\begin{conclusionframe}{RQ4: Enablers and requirements}
The identified enablers of twin transition are primarily of technical nature, with a pronounced need for real-time data analytics and AI methods. Non-technical enablers relate to policy-making and innovation methodology.
\end{conclusionframe}

\subsection{Challenging factors of twin transition (RQ5)}\label{sec:results-rq5}

\tabref{tab:results-rq5} reports the challenges typically faced in twin transition journeys.
We observe the emergence of two main topics: \textit{technical and methodological} challenges, and \textit{business and organizational} ones. Technical challenges slightly outweigh non-technical
ones, encountered in \xofy{15}{24}{} and \xofy{12}{24}{} studies, respectively.
Across these categories, the dominantly recognized challenges are the \textit{complexity} (\xofyp{8}{24}) and \textit{high costs} (\xofyp{7}{24}) of twin transition -- one challenge from the technical and one from the business themes. The perceived complexity of twin transition is related to, e.g., its convoluted socio-technical effects~\cite{makitie2023digital}, and the intricacies of assessing the precise ecological impact of digital technology, such as AI~\cite{chiu2020can}. This complexity has implications for \textit{finding trade-offs} between sustainability and digitalization~\cite{celeste2023digital,bianchini2023environmental,sanchez2024european} and understanding and controlling rebound effects~\cite{celeste2023digital,niet2024framing,pan2024policy}.

Only a few instances, \xofy{3}{24}{} of \textit{technological} challenges are found, related to, e.g., the vulnerabilities introduced by digitization~\cite{muller2023digitisation}, cyber security risks and data protection issues~\cite{pan2024policy}, and the biases of AI~\cite{niet2024framing}.

\begin{table}[htb]
\footnotesize
\centering
\caption{Key challenging factors of twin transition}
\label{tab:results-rq5}
\begin{tabular}{@{}llp{3.5cm}@{}}
\toprule
\multicolumn{1}{c}{\textbf{Domain}} & \multicolumn{1}{c}{\textbf{\#Studies}} & \multicolumn{1}{c}{\textbf{Studies}} \\ \midrule

\textbf{Tech. and methodology} & \maindatabar{15} & \\

\subcategory{}Complexity & \subdatabar{8} & \cite{balanici2023f5g,baum2024will,chiu2020can,korucuk2022assessing,kovacic2024twin,makitie2023digital,sanchez2024european,stucki2024data} \\

\subcategory{}Finding trade-offs& \subdatabar{4} & \cite{bianchini2023environmental,celeste2023digital,kovacic2024twin,sanchez2024european} \\

\subcategory{}Rebound effects & \subdatabar{3} & \cite{celeste2023digital,niet2024framing,pan2024policy} \\

\subcategory{}Technological challenges & \subdatabar{3} & \cite{muller2023digitisation,niet2024framing,pan2024policy} \\

\subcategory{}Other & \subdatabar{4} & \cite{meijer2024perspectives,pan2024policy,sanchez2024european,reich2024towards} \\

\textbf{Businesses and organizations} & \maindatabar{12} & \\
\subcategory{}High costs & \subdatabar{7} & \cite{baum2024will,chen2023enabling,gao2024eus,gigauri2023digital,myyrylainen2023small,pan2024policy,rehman2023twin} \\

\subcategory{}Policies and law & \subdatabar{4} & \cite{brueck2024chinas,kovacic2024twin,meijer2024perspectives,poscic2024role} \\

\subcategory{}No qualified workforce & \subdatabar{3} & \cite{gao2024eus,meijer2024perspectives,myyrylainen2023small} \\

\subcategory{}Other & \subdatabar{3} & \cite{makitie2023digital,myyrylainen2023small,pan2024policy} \\

\bottomrule

\end{tabular}
\end{table}

\begin{conclusionframe}{RQ5: Challenging factors}
The recognized challenges of twin transition are mostly related to its complexity and costs, with little focus on technology-related challenges.
\end{conclusionframe}

\section{Takeaways and Recommendations}\label{sec:discussion}

We now synthesize the results of our empirical inquiry into key takeaways and identify research and development directions in support of twin transition.

\subsection{Terminology issues hint at lingering techno-optimism}

\observation{The vast majority of primary studies use the wrong concept to label their transition model and oversell their contributions (RQ1)}

\xofy{92}{113}{} studies that label themselves as ``twin transition'' are, in fact, not aligned with its definition. In \xofy{42}{113}{} cases, there is no link between the two transition streams, and in \xofy{50}{113}{} cases, sustainability transition is aided by digitalization. That is, in the vast majority of twin transition research, \textbf{the sustainability implications of digitalization are not considered}. This raises the threat of ecological and social implications of digitalization going unnoticed, and digitalization continuing to be looked at as a silver bullet to treat sustainability issues. This \textbf{unwarranted techno-optimism}---i.e., the belief that technology alone can solve environmental and societal problems without fundamentally rethinking the structure of our growth-based economies~
\cite{alexander2019critique}---is well-documented in the state of the art~\cite{bork2024role} and poses a serious problem as ICT is set to become one of the leading industries in carbon footprint generation already in this decade~\cite{belkhir2018assessing,ituict}.
Unfortunately, evidence suggests that Big Tech is actively misshaping public discourse and normalizes greenwashing~\cite{pan2024policy}. This evidence is also present in mainstream media and available to the general public, e.g.,~\cite{aws2024x,dechant2024nearly}, just to cite some recent articles.




\begin{recoframe}{Recommendation 1}
Build sustainable \textit{systems} by sustainable \textit{methods}.
\end{recoframe}

\textcite{pan2024policy} warn that moderate sustainability ambitions and a solely profit-focused mindset hinder the adoption of sustainable technology. Evidently, stronger forms of innovation are needed that have the ability to reconfigure the structure of our socio-technical systems to align with sustainability~\cite{makitie2023digital}. This starts with a change in design philosophy and promoting sustainability to a first principle in technology~\cite{chiu2020can}.
%
Following the framework for circular systems engineering by \textcite{david2024circular}, we recommend the following steps.

\noindent (1) \textit{Develop \textbf{sustainability} assessment techniques for the \textbf{developed system.}} Real-time monitoring to reduce environmental impact~\cite{sanchez2024european} and carbon footprint~\cite{muller2023digitisation}, e.g., through digital twins~\cite{makitie2023digital} or internet of things (IoT)~\cite{rehman2023twin} are some of the methods recommended in the primary studies.

\noindent (2) \textit{Develop \textbf{sustainability} assessment techniques for the \textbf{methods} (digital technology).} Studies call for increasing the ecological transparency of AI~\cite{niet2024framing} and using Digital Product Passports (DPPs) to collect lifecycle data, including sustainability-related information~\cite{reich2024towards}.
We recommend relying on standardized techniques, e.g., the ISO/IEC 21031:2024 Software Carbon Intensity (SCI) specification~\cite{iso21031}
to compare alternatives in terms of carbon emissions.

\noindent (3) \textit{\textbf{Connect} system and method sustainability and \textbf{find trade-offs}.} By understanding the lifecycle model of the developed system, trade-offs between systems and method sustainability can be identified. For example, developing an AI agent to optimize the energy consumption of a smart building will have energy costs in the training phase but might realize more savings in deployment~\cite{david2024circular}.
\textit{What exactly is the right amount of training?} -- is the question to ask.
Finding such trade-offs opens opportunities for \textit{controlled ICT degrowth} strategies in which resource utilization and waste can be scaled back~\cite{david2024on}.

\subsection{The complexity of sustainability challenges twin transition}\label{sec:discussion-collab}

\observation{Sustainability dimensions rarely appear in isolation (RQ2)}

A key observation regarding the sustainability aspects of twin transition (RQ2, \secref{sec:results-rq2}) is that sustainability dimensions hardly ever appear in isolation. 
In \xofy{20}{24}{} studies, at least two sustainability dimensions are identified, with environmental sustainability being the only one that is present in every work (\figref{fig:rq2-sustainability-dims-upset}). This observation suggests the consistent usage of the \textbf{strong notion of sustainability}~\cite{bebbington2001full}, which acknowledges that sustainability dimensions cannot be isolated without substantial threats to validity, mostly due to the multi-systemic nature of sustainability~\cite{becker2016requirements}.
However, the elaborate notion of sustainability comes at the price of \textbf{increased complexity}. The combination of different sustainability dimensions renders the analysis of twin transition goals challenging, and necessitates the involvement of more domains and experts. Accordingly, the leading challenge in twin transition (\tabref{tab:results-rq5}) is its complexity.

\begin{recoframe}{Recommendation 2}
Develop \textit{ontology-enabled collaborative modeling} methods to allow technical experts to coordinate across different sustainability dimensions.
\end{recoframe}

%

Conceptual modeling~\cite{storey2023conceptual} and model-driven engineering (MDE)~\cite{schmidt2006model-driven} are proven methods in the analysis of complex problems. Through the power of abstraction, modeling reduces the accidental complexity~\cite{atkinson2008reducing} of the problem at hand and allows domain experts to focus only on its relevant properties.
The complexity of sustainability in twin transition necessitates the involvement of various expertise and domains, further evidenced by the diversity of technical enablers (\tabref{tab:results-rq5}).
\textbf{Collaborative modeling}~\cite{franzago2018collaborative,david2021collaborative} excels in supporting such heterogeneous settings by allowing multiple experts to cooperate and be aware of each others' work on shared models.
Alas, collaborative modeling (cf. participatory modeling in the next section) is seldom encountered in contemporary design methods for sustainability~\cite{barisic2025modelling,lago2024sustainability}.

To cope with the divergent vocabularies of the involved disciplines (necessitated by the complexity of sustainability ambitions), we recommend researching \textbf{ontology-enabled methods} for collaborative modeling. Ontologies are structured representations of domain knowledge and enable reasoning over multiple domains~\cite{guarino2009what}. In truly multi-paradigm settings, misunderstandings due to the reliance on (unformalized) natural language, misaligned assumptions and inconsistencies occur naturally.
The benefits of ontologies in multi-disciplinary settings has been proven in many domains from cyber-physical systems~\cite{vanherpen2016ontological}, through the development of digital twins~\cite{bao2022ontology-based}, to the analysis of FAIR data principles~\cite{bernasconi2023ontological}.

In design for sustainability, ontologies can be used to capture domain concepts, as well as values of ethics, utility, and justice~\cite{guizzardi2023ontology-based} associated with these concepts. Ontologies can also help formalize preferences and drive collective decision-making. Finally, computer-aided ontological reasoning~\cite{eiter2006reasoning} can support semantic consistency among experts who might not use the same vocabularies. 

\subsection{Non-technical stakeholders and users need to be involved}

\observation{The technical nature of enablers (RQ4) and challenges (RQ5) is not aligned with the mostly non-technical stakeholders (R3)}

\xofy{14}{24}{} studies identify technical enablers and \xofy{15}{24}{} identify technical challenges, while only \xofy{4}{24}{} identify technical stakeholders. This imbalance suggests a mismatch between the technical implementation of twin transition and the stakeholders involved in it.

Bridging this gap necessitates the \textbf{involvement of information systems researchers and practitioners in the early phases of twin transition}, i.e., conceptualization and strategic planning.
At the same time, studies warn that effective twin transition cannot be achieved without the \textbf{involvement of political and social stakeholders}~\cite{sanchez2024european,stucki2024data}---an advice that aligns well with the high number of non-technical stakeholders (present in \xofyp{20}{24} primary studies; RQ3). Without these stakeholders, complex issues like community resistance, economic trade-offs and local impacts might derail costly twin transition efforts~\cite{sanchez2024european}. \textcite{brunori2022agriculture} also warn that \quotetext{digitalization, driven only by market forces and in the absence of an effective policy environment} might render systems far from sustainability.
Finally, to reach higher leverage points~\cite{penzenstadler2018software}, there is a clear need to \textbf{engage with end-users} with the primary intent to change consumer preferences towards sustainable alternatives or to enforce such a behavior~\cite{chiu2020can}, e.g., through design methods such as Poka-yoke (``mistake-proofing'')~\cite{chen2023enabling}. 
While collaborative modeling excels in technical settings (\secref{sec:discussion-collab}), it is less suitable to foster the involvement of non-technical stakeholders.

\begin{recoframe}{Recommendation 3}
Develop \textit{participatory modeling} practices to foster cooperation between information systems researchers and non-technical stakeholders, such as government bodies, social stakeholders, and end-users.
\end{recoframe}


\textbf{Participatory modeling}~\cite{voinov2018tools} (PM) facilitates high-level modeling, e.g., through systems dynamics~\cite{nabavi2017boundary}, in which non-experts involved in modeling.
PM is a proven method in the design for sustainability~\cite{manellanga2024participatory} as its informal modeling process fosters diversity and inclusiveness.
To identify \textbf{useful stakeholders}, \textcite{becker2016requirements} recommend to \quotetext{minimize the number of stakeholders involved, and focus on those who have influence}. In particular, one should \quotetext{focus on internal stakeholders and exclude unreachable stakeholders}~\cite{becker2016requirements}.

\textcite{midgley2014goals} and \textcite{nabavi2017boundary} observe the need for combining PM with a more technical cooperative modeling paradigm in the design of sustainable systems.
Thus, we recommend IS researchers to develop combined participatory-collaborative modeling methods, e.g., by relying on PM in the early phases of cooperation and gradually refining models into more technical and actionable ones via collaborative modeling. Actionable models have been articulated as a clear need in the sampled studies, e.g., by \textcite{stucki2024data}.


\subsection{Informing key stakeholders and aiding decision-making}

\observation{Top-level stakeholders are identified in a large number (RQ3) but no enablers relate directly to information delivery (RQ4)}

Government actors and corporate leaders are identified in a total of \xofy{18}{24}{} primary studies as key stakeholders who have to make decisions in twin transition journeys (\tabref{tab:results-rq3-stakeholder-sectors}, RQ3). Corroborating this finding, a recent report by Gartner predicts that 80\% of CIOs' key performance metrics will be tied to the sustainability of their IT organization by 2027~\cite{gartner2023top}.
%
However, knowledge about integrating sustainability principles into corporate digital strategies is not present in nowadays' prototypical CIOs' profiles. In addition, assessing the precise impact of digital technologies like AI, blockchain, data centers, and big data with its ecosystem is a complex task.
Thus, there is a clear need for \textbf{new methods and assessment frameworks to inform and guide stakeholders} in evaluating the impact of digital technologies on sustainability. Key reported challenges include assessing the precise impact of digital technology, e.g., AI~\cite{niet2024framing} and its ecosystem~\cite{chiu2020can}; and the complexity of finding sustainability trade-offs~\cite{kovacic2024twin}.

\begin{recoframe}{Recommendation 4}
Implement real-time data-enabled information delivery and decision-support, such as applied observability, also drawing on infonomics.
\end{recoframe}



Studies often call for real-time data collection to inform stakeholders. Some of the identified benefits of real-time monitoring include reduced environmental impact~\cite{sanchez2024european}, reduced waste~\cite{chen2023enabling}, and increased efficiency~\cite{chen2023enabling}. However, these improvements mostly benefit operators, not decision-makers.
To turn data into valuable \textbf{information} for decision-makers, we recommend drawing on infonomics, the discipline of asserting economic value to information~\cite{laney2017infonomics}. We recommend relying on established information valuation models~\cite{laney2017infonomics,david2024infonomics} as they provide insights into how to improve data collection mechanisms for higher information value.

\textbf{Applied observability}~\cite{gartner2022applied} may be of value when systems are not directly observable,
e.g., the sustainability properties of complex eco-socio-technical systems. 
Applied observability is the practice of inferring states of a system from the data it generates, and ensuring that data is available across departments and applications to enable real-time decision-making~\cite{gartner2022applied}.



\textbf{Modeling and simulation} are additional enablers, e.g., for forecasting carbon emission of economic activities~\cite{sanchez2024european}. As pointed out by \textcite{stucki2024data}, quantitative (e.g., numerical) models are more actionable and useful in supporting sustainability decisions and tracking KPIs than qualitative ones. Coupled with digital twin technology~\cite{makitie2023digital}, modeling and simulation can drive applied observability throughout the lifecycle of the system and across organizations.

\subsection{Improving social sustainability through HCI}

\observation{Social sustainability is not emphasized in twin transition and appears only in combination with environmental sustainability (RQ1)}

This observation is also demonstrated in the commonly used definition of twin transition ``greening of and by IT and data''~\cite{blum2022what}, narrowing the focus to environmental sustainability, leaving little room for social concerns to be incorporated. 

%
%

Within the context of the individual, social sustainability of digital technology is characterized by properties such as accessibility, trust, and human-centered design~\cite{dix2003human-computer}.
%
%
%
%
Ignoring social sustainability leads to
unwanted consequences, such as unskilled workforce, aversion of digital technology~\cite{niet2024framing} (in the most pertinent and timely example: AI anxiety~\cite{johnson2017ai}). Clearly, there is a need for narrowing the gap between humans and digital technology; ideally, through systematic methods.
\textbf{Human-computer interaction (HCI)} is a prime candidate to serve as such a method. HCI prioritizes improving human experience by situating the humans in the loop~\cite{ehsan2020humancentered} and through that, fostering trust in systems~\cite{brockhoff2021process}.
Notable instances of human-in-the-loop techniques include improved and active user involvement, directed user focus~\cite{gulliksen2003key}, improved explainability and transparency~\cite{ehsan2020humancentered}, and accessibility and inclusion~\cite{abascal2007fundamentals}. Successful twin transition should improve in social sustainability by implementing such techniques.

\begin{recoframe}{Recommendation 5}
Utilize HCI principles as a general framework to enhance social sustainability of digitalization in twin transition.
\end{recoframe}

There are some known examples of HCI succeeding at improving the social sustainability of digital technology.

\textbf{Explainable AI}, the technique of creating more transparent and, ultimately, explainable machine learning models~\cite{al-ansari2024user-centered}, is a notable example that aims to improve the social sustainability of digital technology.
The limited transparency of AI often results in issues such as AI anxiety~\cite{johnson2017ai} and reluctance to use AI~\cite{niet2024framing}, which eventually limit the utility of AI applications.
Explainable AI fosters trust by making AI models more transparent, and allows stakeholders to hold the system accountable for its decisions~\cite{berger2021back}.
Specifically, improving the \textbf{ecological transparency of AI} can be a contributor to the success of twin transition~\cite{niet2024framing}. Such directions are apt responses to the likely lacking sustainability-related knowledge in CIOs' profiles.

In modern manufacturing applications, \textbf{digital twin cockpits} improve HCI by visualizing data and allowing interactions with digital twin services through graphical user interfaces (GUI)~\cite{bano2022process-aware}. By that, digital twin cockpits enable users with varying levels of technical knowledge to interact with computers effectively, which is a key HCI principle~\cite{oregan2018graphical}.
\section{Conclusion}\label{sec:conclusion}

In this paper, we reported on our study on \textit{twin transition}, the paradigm of promoting sustainability \textit{by and of} digital technology. Our study shows that twin transition is often misunderstood, limiting its upside and giving rise to unwarranted techno-optimism, while threatening with a widening social divide across regions and sectors.
We believe that researchers have a lot to offer at the current maturity of twin transition. To support this involvement, we derive key recommendations for researchers and practitioners in computing-related fields, including information systems and software engineering.

In future work, we will derive a maturity framework for twin transition to assess companies' capabilities to conduct successful twin transition projects, and validate it through industry cases and expert feedback.
\section*{Acknowledgment}
We acknowledge the support of the Natural Sciences and Engineering Research Council of Canada (NSERC), DGECR-2024-00293 (End-to-end Sustainable Systems Engineering).

\newrefcontext[labelprefix=P]
\printbibliography[keyword={primary},title={Primary studies},resetnumbers=true]

\newrefcontext
\printbibliography[notkeyword={primary},resetnumbers=true]

\end{document}